# Constrained Non-Linear Phase Retrieval for Single Distance X-ray Phase Contrast Tomography

*K. Aditya Mohan\*, Dilworth Y. Parkinson†, Jefferson A. Cuadra\*; \*Lawrence Livermore National Laboratory, Livermore, CA USA;
†Lawrence Berkeley National Laboratory, Berkeley, CA USA;*

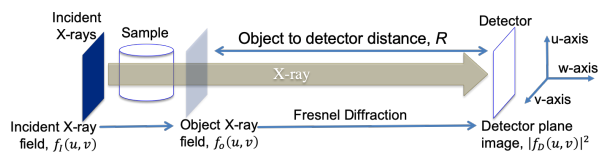

**Figure 1.** *System diagram of XPCT. In XPCT, the propagation distance from object to detector is increased to allow the phase variations in the X-ray field exiting the object to develop into intensity variations called phase contrast at the detector.*

## Abstract

*X-ray phase contrast tomography (XPCT) is widely used for 3D imaging of objects with weak contrast in X-ray absorption index but strong contrast in refractive index decrement. To reconstruct an object imaged using XPCT, phase retrieval algorithms are first used to estimate the X-ray phase projections, which is the 2D projection of the refractive index decrement, at each view. Phase retrieval is followed by refractive index decrement reconstruction from the phase projections using an algorithm such as filtered back projection (FBP). In practice, phase retrieval is most commonly solved by approximating it as a linear inverse problem. However, this linear approximation often results in artifacts and blurring when the conditions for the approximation are violated.*

*In this paper, we formulate phase retrieval as a non-linear inverse problem, where we solve for the transmission function, which is the negative exponential of the projections, from XPCT measurements. We use a constraint to enforce proportionality between phase and absorption projections. We do not use constraints such as large Fresnel number, slowly varying phase, or Born/Rytov approximations. Our approach also does not require any regularization parameter tuning since there is no explicit sparsity enforcing regularization function. We validate the performance of our non-linear phase retrieval (NLPR) method using both simulated and real synchrotron datasets. We compare NLPR with a popular linear phase retrieval (LPR) approach and show that NLPR achieves sharper reconstructions with higher quantitative accuracy.*

## Introduction

Traditional X-ray computed tomography (XCT) relies on X-ray absorption contrast for imaging. However, many samples, e.g. animal soft tissue in biological imaging and material interfaces with similar atomic number elements, have low absorption contrast. To solve this problem, phase contrast, which is typically many orders of magnitude larger than absorption contrast, is induced by increasing the propagation distance from object to detector [1, 2] (Fig. 1). This technique where tomographic measurements with both absorption and phase contrast are acquired at large object-to-detector propagation distances is called X-ray phase contrast imaging (XPCT).

In XPCT, phase contrast in the measurements is related to the X-ray phase, which is the 2D projection of the refractive index decrement values of the object. In contrast, X-ray absorption is quantified as the projection of the object's absorption index. However, the non-linear dependence of X-ray measurements on the absorption and phase projections makes it challenging to solve the inverse problem of object reconstruction. Typically, a phase retrieval algorithm is first used to estimate the phase projections at each tomographic view. Then, the refractive index decrement values in 3D are reconstructed using an algorithm such as filtered back projection (FBP) [16].

Several phase retrieval algorithms have been proposed for both single distance and multi-distance XPCT [2–11]. These methods either estimate the transmission function, which is the negative exponential of the projections, or directly solve for the projections. Among these methods, the most widely used approaches for single distance phase retrieval use linear analytical closed form solutions to either recover the transmission function or the projections [3–6]. A comparison between the various linear phase retrieval approaches for single distance XPCT is in [2]. Recently, non-linear approaches to single distance phase retrieval have been presented in [1, 10]. These non-linear approaches claim substantial improvements over linear approaches using non-linear forward models and sparsity enforcing regularization functions. However, various reasons including the need for regularization parameter tuning and computational complexity have prevented the widespread use of these methods.

In this paper, we present a novel phase retrieval method that uses a non-linear forward model and a constraint to enforce a linear relation between the absorption and phase projections. The linear constraint can either be used to enforce zero values for the absorption projection or enforce proportionality between absorption and phase projections. The non-linear forward model uses Fresnel transform to express the X-ray measurements at the detector plane as a function of the transmission function. Our forward model does not make any approximations made by linear phase retrieval approaches such as large Fresnel number, slowly varying phase, or Born/Rytov approximations [2].

Our non-linear phase retrieval (NLPR) approach uses numerical optimization to estimate the transmission function such that the XPCT measurements best match its predictions from the non-linear forward model. Since absorption and phase are projection space quantities that are related non-linearly to the trans-mission function, we propose a novel formulation in transmission space that automatically enforces a linear relation between absorption and phase projections during optimization. Since our approach does not use any sparsity enforcing regularization function to model prior information, significant manual effort and time often spent in tuning the regularization parameter is saved. Furthermore, since phase retrieval at each view can be run independently, our approach is easily parallelized on a multi-core processor or high performance computing (HPC) systems.

## Measurement Physics

The physics of X-ray interaction with an object is defined in terms of the object's 3D variation in absorption index, $\beta(u,v,w)$, and refractive index decrement, $\delta(u,v,w)$, where $(u,v,w)$ represent the 3D Cartesian coordinates. The absorption index, $\beta(u,v,w)$, modulates the amplitude (intensity) of the incoming X-ray field and the refractive index decrement, $\delta(u,v,w)$, modulates the X-ray phase. After propagation through the object, the effective change in X-ray intensity is quantified by the absorption projection,

$$A(u,v) = \frac{2\pi}{\lambda} \int_w \beta(u,v,w) dw, \quad (1)$$

and the change in X-ray phase is given by the phase projection,

$$\phi(u,v) = \frac{2\pi}{\lambda} \int_w \delta(u,v,w) dw, \quad (2)$$

where X-ray propagation is assumed to be along the $w-$axis and $\lambda$ is the X-ray wavelength.

Next, we will express the X-ray field exiting the object, $f_O(u,v)$, as a function of the X-ray field incident on the object, $f_I(u,v)$ (Fig. 1). The transformation from $f_I(u,v)$ to $f_O(u,v)$ is dependent on the transmission function $T(u,v)$ defined as,

$$T(u,v) = \exp(-A(u,v) - i\phi(u,v)), \quad (3)$$

where $i = \sqrt{-1}$, $T(u,v) \in \mathscr{C}$ is complex valued, and $A(u,v) \in \mathscr{R}$ and $\phi(u,v) \in \mathscr{R}$ are real valued. Then, the X-ray field $f_O(u,v)$ exiting the object is given by,

$$f_O(u,v) = f_I(u,v) T(u,v). \quad (4)$$

The X-ray field $f_O(u,v)$ undergoes Fresnel diffraction as it propagates towards the detector. If $F_D(\mu,\nu)$ and $F_O(\mu,\nu)$ denote the Fourier transforms of $f_D(u,v)$ and $f_O(u,v)$, respectively, then,

$$F_D(\mu,\nu) = F_O(\mu,\nu) \exp\left(-i\pi\lambda R \left(\mu^2 + \nu^2\right)\right), \quad (5)$$

where we have ignored constant phase terms, $R$ is the object-to-detector distance, and $(\mu,\nu)$ are the 2D Fourier frequency coordinates. However, each detector pixel is only able to measure the intensity of the incoming X-ray field. Hence, the measurement at detector pixel $(j,k)$ is modeled as,

$$\tilde{y}(j,k) = |f_D(j\Delta, k\Delta)|^2, \quad (6)$$

where $\Delta$ is the pixel width of the detector and $|\cdot|$ denotes magnitude.

In our application, the phase and magnitude of the incident X-ray field $f_I(u,v)$ is not known. The magnitude of $f_I(u,v)$ could be estimated if the bright-field (a.k.a. flat-field) measurements were made at the object's plane of X-ray incidence. However, the bright-field measurements are made when the object-to-detector distance is $R$, which is the distance optimized for phase contrast. Hence, we make some simplifying assumptions on $f_I(u,v)$ that also simplifies the subsequent non-linear phase retrieval method. First, we assume that the incident X-ray field $f_I(u,v)$ is a plane wave with constant phase. Hence, without loss of generality, we assume $f_I(u,v)$ has zero phase since any information on constant phase terms are lost when the detector makes intensity measurements. We also make an approximation where the effect of the magnitude of $f_I(u,v)$ is ignored by normalizing $\tilde{y}(j,k)$ with its bright and dark fields. Let $b(j,k)$ denote the bright-field measurements acquired by the detector in the absence of the object at a object-to-detector distance of $R$ and $d(j,k)$ denote the dark field measurements acquired in the absence of X-rays. Let $N_u$ and $N_v$ denote the number of detector pixels along the $u-$axis and $v-$axis, respectively. Then, let the square root of the normalized detector image be given by,

$$y(j,k) = \sqrt{\frac{\tilde{y}(j,k) - d(j,k)}{b(j,k) - d(j,k)}}, \quad (7)$$

where $1 \leq j \leq N_u$ and $1 \leq k \leq N_v$.

## Forward Model

In this section, we derive a forward model in discrete space that expresses the square root measurements $y(j,k)$ in terms of the transmission function. Let the discrete space equivalent of the transmission function $T(u,v)$ be denoted by $\tilde{x}(j,k)$. However, there is no unique solution to the problem of estimating the complex valued $\tilde{x}(j,k)$ from the real valued $y(j,k)$. To solve the uniqueness problem, we constrain the feasible solution space for the transmission function $\tilde{x}(j,k)$. In particular, we assume that the complex valued $\tilde{x}(j,k)$ can be represented in terms of a real valued $x(j,k)$ such that,

$$\tilde{x}(j,k) = x^{\alpha + i\gamma}(j,k), \quad (8)$$

where $\alpha$ and $\gamma$ are real valued constants used to constrain the domain space of $\tilde{x}(j,k)$. Thus, the problem reduces to estimation of the real valued $x(j,k)$ from $y(j,k)$.

For the forward model, we use the discrete frequency equivalent of equation (5). Let $z(j,k)$ be the discrete space equivalent of the continuous space X-ray field $f_D(u,v)$ at the detector plane. Let the discrete Fourier transforms (DFT) of $z(j,k)$ and $\tilde{x}(j,k)$ be denoted by $Z(p,q)$ and $\tilde{X}(p,q)$, respectively, where $(p,q)$ are DFT coordinates. Then,

$$Z(p,q) = \tilde{X}(p,q) H(p,q), \quad (9)$$

where $H(p,q)$ is the Fresnel transform in discrete Fourier space given by,

$$H(p,q) = \exp\left(-i\pi\lambda R \left(p^2 \Delta_\mu^2 + q^2 \Delta_\nu^2\right)\right). \quad (10)$$

Here, $\Delta_\mu = \frac{1}{N_u \Delta}$ and $\Delta_\nu = \frac{1}{N_v \Delta}$ are the widths of the Fourier frequency bins along the $\mu$ and $\nu$ frequency axes (equation (5)), respectively. Then, $z(j,k)$ is obtained by inverse discrete Fourier transform (IDFT) of $Z(p,q)$. Thus, the forward model in discrete space is,

$$y(j,k) = |z(j,k)| + w(j,k), \qquad (11)$$

where $w(j,k)$ is additive Gaussian noise.

For mathematical convenience, we will express the forward model in vector form. Let $\mathbf{y}$, $\mathbf{z}$, $\mathbf{x}$, and $\tilde{\mathbf{x}}$ be vectors in raster order corresponding to $y(j,k)$, $z(j,k)$, $x(j,k)$, and $\tilde{x}(j,k)$, respectively. The Fresnel transform in equation (9) including the DFT and IDFT operations are coded into a matrix $\mathbf{H}$ such that $\mathbf{z} = \mathbf{H}\tilde{\mathbf{x}}$. Then, the forward model that expresses the measurement $\mathbf{y}$ in terms of $\tilde{\mathbf{x}} = \mathbf{x}^{\alpha + i\gamma}$ is given by,

$$\mathbf{y} = \left| \mathbf{H}\mathbf{x}^{\alpha + i\gamma} \right| + \mathbf{w}. \qquad (12)$$

where $|\cdot|$ denotes element-wise magnitude and $\mathbf{w}$ is noise vector consisting of all $w(j,k)$ in raster order.

## Phase Retrieval
### Linear Phase Retrieval (LPR)

As a baseline, we use the linear phase retrieval (LPR) method presented in Paganin et al. [4]. This method assumes a homogeneous object and solves for the transmission function as a linear analytical transformation of the normalized measurements. It assumes data acquisition at large Fresnel numbers [2] which causes excessive smoothing and artifacts at large object-to-detector distances. The homogeneous assumption for the object is equivalent to assuming that the ratio of the refractive index decrement and the absorption index is a constant given by $\frac{\delta}{\beta}$. In essence, this phase retrieval method is a low pass filter where the amount of smoothing increases with the object-to-detector distance, $R$, and the ratio $\frac{\delta}{\beta}$.

### Non-Linear Phase Retrieval (NLPR)

Our approach to non-linear phase retrieval (NLPR) is to solve for $\mathbf{x}$ from square root measurements $\mathbf{y}$ using the non-linear forward model in equation (12). Thus, NLPR is formulated as,

$$\hat{\mathbf{x}} = \arg\min_{\mathbf{x}} \left\| \mathbf{y} - \left| \mathbf{H}\mathbf{x}^{\alpha + i\gamma} \right| \right\|^2, \qquad (13)$$

where $\mathbf{x}_j \in \mathcal{R}$, $\mathbf{H}_{j,k} \in \mathcal{C}$, $\mathbf{y}_j \in \mathcal{R}$, $|\cdot|$ denotes element-wise magnitude, and $\|\cdot\|$ computes vector magnitude. Note that $\mathbf{x}_j$ and $\mathbf{y}_j$ denote the $j^{th}$ element of the vectors $\mathbf{x}$ and $\mathbf{y}$, respectively. Similarly, $\mathbf{H}_{j,k}$ denotes the element along the $j^{th}$ row and $k^{th}$ column of the matrix $\mathbf{H}$.

To solve the uniqueness problem during phase retrieval, we impose constraints to restrict the feasible solution space when optimizing for the transmission function. In particular, we use a constraint that forces the phase projection to be proportional to the absorption projection [2]. The phase proportional to absorption constraint equivalently assumes that the value at every point inside the object has the same ratio of the refractive index decrement, $\delta$, and the absorption index, $\beta$. The phase proportional to absorption constraint is obtained by setting $\alpha = 1$ and $\gamma = \frac{\delta}{\beta}$. Alternatively, the zero absorption constraint can also be used by a suitable setting of $\alpha$ and $\gamma$.

To solve equation (13), we use the limited memory Broyden–Fletcher–Goldfarb–Shanno (LBFGS) algorithm with positivity constraints [18, 19]. Since LBFGS uses the gradient information for optimization, we derive the gradients of the objective function in equation (13) using algorithmic differentiation and the derivative tables in [12]. We use the LBFGS implementation in the open source software library NLopt [17]. The estimated result from LPR is used as an initial estimate when solving equation (13) using LBFGS.

Once $\hat{\mathbf{x}}$ is computed, the next step is to perform tomographic reconstruction of the refractive index decrement of the object. Note that $\hat{\mathbf{x}}$ is a vector containing the estimates $\hat{x}(j,k)$ in raster order. From equation (2), we see that the phase projections are related to the line integral of the refractive index decrement. Based on equations (3) and (8), the phase projections in discrete space are computed as $\gamma(-\log(\hat{x}(j,k)))$. Once the phase projections are computed at each view, the refractive index decrement is reconstructed using filtered back projection (FBP).

## Results
### Simulated Data

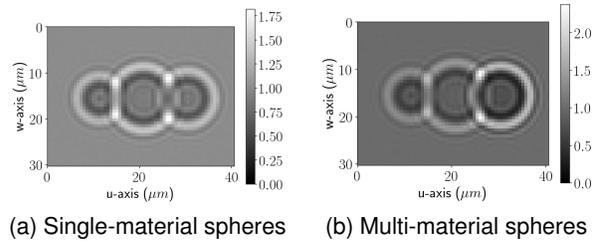

(a) Single-material spheres  (b) Multi-material spheres

***Figure 2.*** *Simulated normalized measurements at first view angle. (a) and (b) show measurements for single material spheres phantom and multi-material spheres phantom, respectively.*

In this section, we use simulated data to validate the performance of NLPR. We simulate a single material phantom and a multi-material phantom each containing three spheres of radii $4\mu m$, $6\mu m$, and $5\mu m$, respectively[1]. The spheres in the single material phantom have the same refractive index decrement, $\delta$, of $1.67 \times 10^{-6}$ and absorption index, $\beta$, of $4.77 \times 10^{-9}$. These $\delta$ and $\beta$ values correspond to the compound Silicon Carbide (SiC) at a X-ray energy of $20 keV$. The three spheres of the multi-material phantom have absorption index values of $4.77 \times 10^{-8}$ ($10\times$ larger), $4.77 \times 10^{-9}$, and $4.77 \times 10^{-9}$, respectively, while the refractive index decrement values are $1.67 \times 10^{-6}$, $1.67 \times 10^{-6}$, and $3.34 \times 10^{-6}$ ($2\times$ larger), respectively.

The shape of the ground-truth volume containing the spheres was $96 \times 128 \times 128$ with a voxel width of $0.3225\mu m$. From the ground-truth volume, we used equation (12) to simulate X-ray images (radiographs) of size $48 \times 64$ at a pixel width of $0.645\mu m$. Radiographs were simulated with Poisson noise at 64 different views equally spaced over an angular range of $180^0$. The object-to-detector distance was $100 mm$ and X-ray energy was $20 keV$. Before any comparisons with the proposed method, the ground-truth volume was downsampled to a size of $48 \times 64 \times 64$. The simulated measurements from the single material and multi-material phantoms at the first view angle are shown in Fig. 2. Before running LPR and NLPR, each X-ray image was padded to 1.5× the original size in each dimension using edge padding. For the LBFGS optimization in NLPR, we used a value of $10^{-6}$ for the relative threshold on the optimization parameters, **x**, in NLopt [17]. In equation (13), we used $\alpha = 1$ and $\gamma = 350$.

---
[1]From left to right in each image of Fig. 2 and Fig. 3

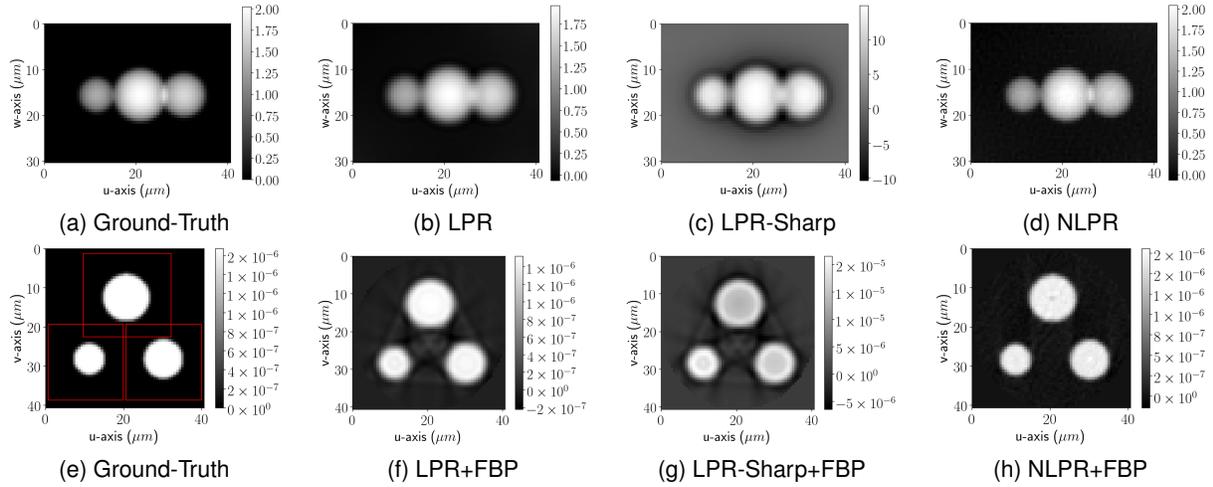

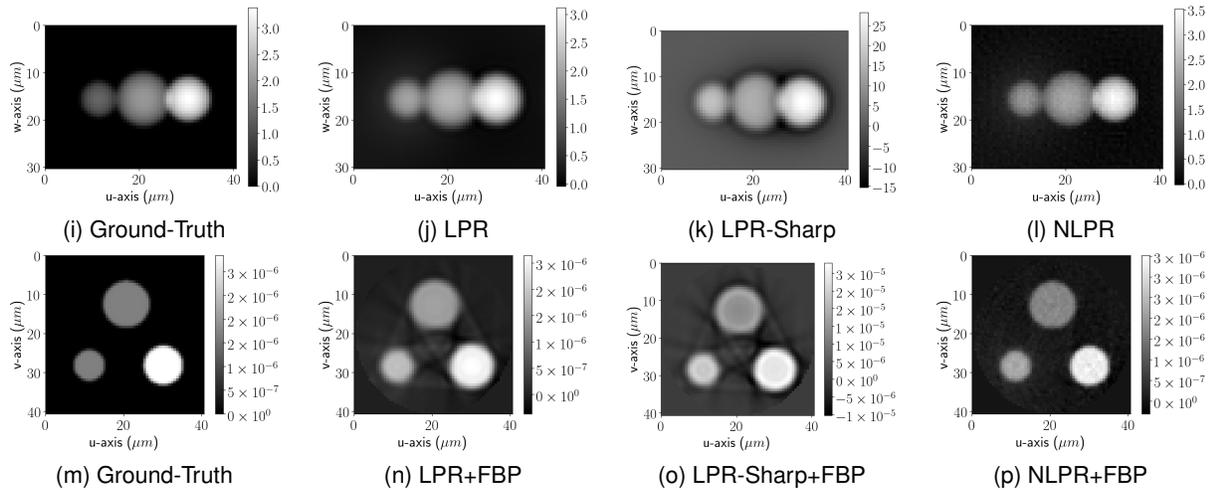

*Figure 3.* Simulated data reconstruction comparison. (a-d) and (i-l) show the ground-truth and retrieved phases using various methods. (e-h) and (m-p) show the ground-truth and reconstructed refractive index decrement using various methods. (a-h) and (i-p) compares results for single material spheres data and multi-material spheres data, respectively. LPR refers to the method in Paganin et al. [4] that uses the true value of $R = 100mm$. LPR-Sharp is the method in Paganin et al. [4] but assuming an incorrect lower value of $R = 5mm$ for improved sharpness. LPR results in excessive smoothing and streak artifacts as shown in (f,g,n,o) while NLPR produces an accurate artifact-free reconstruction of the object with sharp edges.

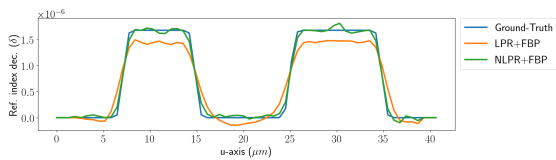

*Figure 4.* Line profile comparison of the reconstructed refractive index decrement through the center of the bottom two circles (spherical cross-sections) in Fig. 3 (e,f,h).

We do a performance comparison between the conventional LPR and the newly proposed NLPR in Fig. 3, Fig. 4, and Fig. 5. LPR refers to the linear phase retrieval approach presented in Paganin et al. [4] that is used with the correct object-to-detector distance value of $R = 100mm$. Alternatively, LPR-Sharp achieves sharper images by using the approach in Paganin et al. [4] but with the lower and incorrect object-to-detector distance of $R = 5mm$. From Fig. 3 (e-h) and Fig. 3 (m-p), we can see that LPR results in blurred reconstructions with streak artifacts while NLPR produces accurate reconstructions without any artifacts. Also, a drastic lowering of $R$ from its true value of $100mm$ still resulted in

**Quantifying performance for single material spheres**

| GT | LPR+FBP | LPR-Sharp+FBP | NLPR+FBP |
|---|---|---|---|
| $51.2\mu m^2$ | $58.7\mu m^2$ | $56.6\mu m^2$ | $51.6\mu m^2$ |
| $114.4\mu m^2$ | $127.7\mu m^2$ | $127.3\mu m^2$ | $114.0\mu m^2$ |
| $80.3\mu m^2$ | $87.4\mu m^2$ | $86.9\mu m^2$ | $80.3\mu m^2$ |

(a) Areas of circles (spherical cross-sections) in Fig. 3 (e-h)

| | LPR+FBP | LPR-Sharp+FBP | NLPR+FBP |
|---|---|---|---|
| Phase | $9.95 \times 10^{-2}$ | 5.36 | $4.98 \times 10^{-2}$ |
| Ref. In. Dec. | $7.31 \times 10^{-8}$ | $2.99 \times 10^{-6}$ | $3.61 \times 10^{-8}$ |

(b) RMSE computed over whole reconstructed volume

**Quantifying performance for multi-material spheres**

| GT | LPR+FBP | LPR-Sharp+FBP | NLPR+FBP |
|---|---|---|---|
| $51.2\mu m^2$ | $61.6\mu m^2$ | $56.6\mu m^2$ | $51.2\mu m^2$ |
| $114.4\mu m^2$ | $127.3\mu m^2$ | $126.9\mu m^2$ | $113.2\mu m^2$ |
| $80.3\mu m^2$ | $103.2\mu m^2$ | $99.8\mu m^2$ | $80.7\mu m^2$ |

(c) Areas of circles (spherical cross-sections) in Fig. 3 (m-p)

| LPR+FBP | LPR-Sharp+FBP | NLPR+FBP |
|---|---|---|
| $1.98 \times 10^{-7}$ | $2.02 \times 10^{-5}$ | $4.23 \times 10^{-7}$ |
| $1.94 \times 10^{-7}$ | $1.34 \times 10^{-5}$ | $1.23 \times 10^{-7}$ |
| $7.58 \times 10^{-7}$ | $2.62 \times 10^{-5}$ | $2.43 \times 10^{-7}$ |

(d) RMSE within the region of each circle in Fig. 3 (n-p)

***Figure 5.*** *(a,b) and (c,d) quantifies the accuracy and sharpness of the refractive index decrement reconstruction for the single material spheres data and multi-material spheres data, respectively. (a) and (c) gives the areas for the three circles in Fig. 3 (e-h) and Fig. 3 (m-p), respectively. The areas with NLPR closely matches the ground-truth (GT) while LPR and LPR-Sharp overestimate the area. (b) computes the root mean squared error (RMSE) over the whole volume for single material spheres reconstruction. (d) computes the RMSE within each circle for the multi-material reconstructed images in Fig. 3 (n-p).*

reconstructions with significant amount of blur that also contain phase contrast fringes as shown in Fig. 3 (g,o). From Fig. 3 (m,p), we can see that NLPR accurately reconstructs the shape in spite of the presence of materials with different and varying values for $\frac{\delta}{\beta}$. In Fig. 3 (n-p), while it was assumed that $\frac{\delta}{\beta} = 350$, the true values of $\frac{\delta}{\beta}$ for the three spheres were 35, 350, and 700.

We also observed that the smoothing nature of LPR overestimates the area of region occupied by the spheres. To demonstrate this effect, we segment each circle in Fig. 3 (e-h,m-p), compute the area of each circle, and compare the computed areas in Fig. 5 (a,c). The region within each red bounding box as shown in Fig. 3 (e) is extracted and segmented using Otsu thresholding [14]. After thresholding, each circular area is computed by summing the number of positive pixels and multiplying by the square of the pixel width. From Fig. 5 (a,c), we can see that the areas with NLPR almost exactly matches the ground-truth while the areas are severely overestimated when using LPR. For the single material data, the root mean squared error (RMSE) in Fig. 5 (b) between the reconstructed and ground-truth values over the whole volume shows that NLPR has higher quantitative accuracy than LPR. We also note that LPR-Sharp results in severely degraded quantitative accuracy due to the incorrect assumption for the object-to-detector distance $R$. In Fig. 5 (d), RMSE estimates computed within the region of each circle in Fig. 3 (n-p) is presented for multi-material data. We can see that LPR has a lower RMSE than NLPR for the first sphere with $\frac{\delta}{\beta} = 35$, while NLPR has lower RMSE for the other two spheres.

## Experimental Data

In order to validate NLPR with experimental data, we used a synchrotron to perform X-ray phase contrast imaging of an object consisting of SiC fibers. Radiographs were acquired at 256 different views equally spaced over an angular range of 180 degrees. The shape of each radiograph is $320 \times 324$ with a pixel width of $0.645\mu m$. The X-ray energy was $20keV$ and the object-to-detector distance was fixed at $R = 100mm$ which ensured significant amount of phase contrast in the measured images as shown in Fig. 6 (a,b). Fig. 6 (a) shows the normalized measurements at the first view angle and Fig. 6 (b) shows the sinogram along the center row of Fig. 6 (a) over all the view angles. Before running LPR and NLPR, each X-ray image was padded to $1.5\times$ the original size in each dimension using edge padding. For the LBFGS optimization in NLPR, we used a value of $10^{-5}$ for the relative threshold on the optimization parameters, **x**, in NLopt [17]. In equation (13), we used $\alpha = 1$ and $\gamma = 350$.

The reconstructed phase at the first view angle using NLPR and LPR are shown in Fig. 6 (c,d). The refractive index decrement is reconstructed in 3D from the phases using FBP and compared in Fig. 7 and Fig. 8. From Fig. 6, we can see that the SiC fibers are localized to a central region that is smaller than the total field of view. To facilitate comparison, we crop the cross-section images of the FBP reconstruction to the region containing the fibers and show them in Fig. 7 (a,b) and Fig. 8 (a,b).

To compare the sharpness of image features, Fig. 7 (c,d) zooms into the circular cross-section of a fiber within the red rectangular box towards the left side of the images in Fig. 7 (a,b).

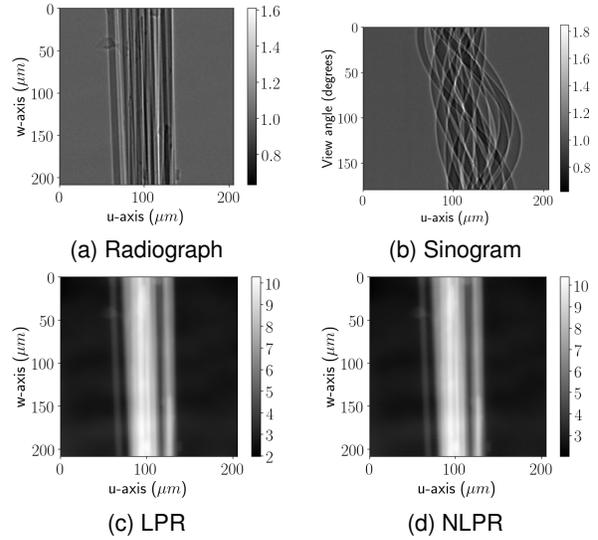

***Figure 6.*** *Experimental data measurements and reconstructed phases. (a) shows the measurements at the first view angle. (b) shows the sinogram over all the views along the center row of (a). (c) and (d) show the retrieved phase using LPR and NLPR, respectively.*

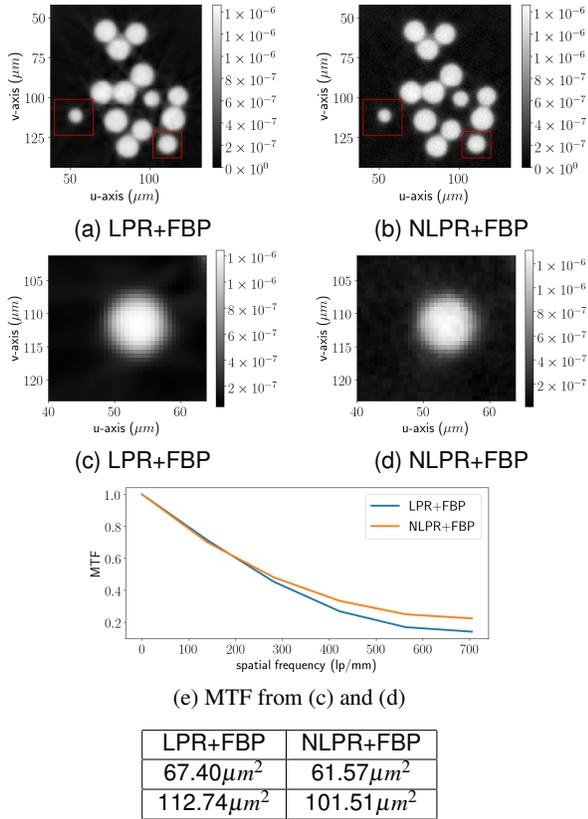

(a) LPR+FBP  (b) NLPR+FBP

(c) LPR+FBP  (d) NLPR+FBP

(e) MTF from (c) and (d)

| LPR+FBP | NLPR+FBP |
|---|---|
| $67.40\mu m^2$ | $61.57\mu m^2$ |
| $112.74\mu m^2$ | $101.51\mu m^2$ |

(f) Areas of circles within the red boxes in (a) and (b).

**Figure 7.** Comparison between FBP reconstructed images from LPR and NLPR phase projections. (a) and (b) show reconstructed image slices corresponding to the center row of Fig. 6 (a) and the sinogram in Fig. 6 (b). (c) and (d) zooms and compares the circle bounded by a red rectangle on the left side of (a) and (b), respectively. (e) compares the modulation transfer function (MTF) computed from the circles in (c) and (d). (f) compares the areas of the circles bounded by the two red rectangles in (a) and (b).

The modulation transfer function (MTF) of the circles in Fig. 7 (c,d) are computed using the method in [15] and shown in Fig. 7 (e). We can see that the MTF with NLPR lies above the MTF with LPR, which indicates sharper images with higher resolution when using NLPR. We employ the same approach used to produce Fig. 5 (a,c) and use it to compute the areas of the circular regions within the red bounding boxes in Fig. 7 (a,b). We can see from the computed areas in Fig. 7 (f) that the areas when using NLPR with FBP is lower than the areas when using LPR with FBP. Furthermore, the areas in Fig. 7 (f) can be compared with the areas computed for simulated data in Fig. 5 (a). From Fig. 5 (a), we have reason to conclude that LPR has overestimated the areas of the circular regions in Fig. 7 (f).

While our phase retrieval algorithms assumed only one material with a constant $\delta/\beta$, we see that there is a second material that is used to hold the fibers in place as shown within the red rectangular box in Fig. 8 (a,b). Fig. 8 (c,d) zooms into the red rectangular region in Fig. 8 (a,b) and clearly shows the sharpness improvement when using NLPR. Thus, the desirable properties of NLPR also apply to materials that do not have the same specified value of $\delta/\beta$ within the sample.

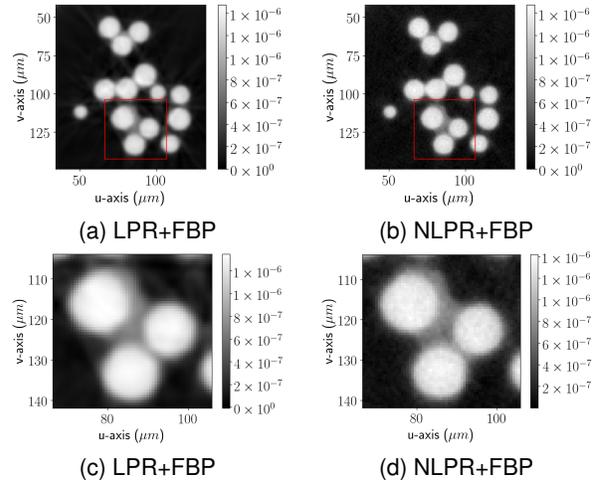

(a) LPR+FBP  (b) NLPR+FBP

(c) LPR+FBP  (d) NLPR+FBP

**Figure 8.** Comparison between FBP reconstructed images from LPR and NLPR phase projections. (a) and (b) compares the $40^{th}$ cross-section slice from the refractive index decrement reconstructions. (c) and (d) zooms into the area within the red rectangular region of (a) and (b).

## Conclusions

We presented a non-linear phase retrieval (NLPR) method and validated its performance using both simulated and experimental data. We compared NLPR with a popular linear phase retrieval (LPR) approach presented in Paganin et al. [4]. In our experiments, we showed that NLPR resulted in more quantitatively accurate reconstructions than LPR. We also showed that LPR significantly overestimated the object's area due to blurring while NLPR accurately estimated the object's area by reducing the blur.

## Acknowledgments

LLNL-PROC-803597. This work was performed under the auspices of the U.S. Department of Energy by Lawrence Livermore National Laboratory under contract DEAC52- 07NA27344. Lawrence Livermore National Security, LLC. LDRD Funding with tracking number 19-ERD-022 was used on this project. This research used resources of the Advanced Light Source, a DOE Office of Science User Facility under contract no. DE-AC02-05CH11231. We thank Jeff Kallman, Jean-Baptiste Forien, and Kyle Champley from LLNL for useful discussions.

or favoring by the United States government or Lawrence Livermore National Security, LLC. The views and opinions of authors expressed herein do not necessarily state or reflect those of the United States government or Lawrence Livermore National Security, LLC, and shall not be used for advertising or product endorsement purposes. The United States Government retains and the publisher, by accepting the article for publication, acknowledges that the United States Government retains a non-exclusive, paid-up, irrevocable, world-wide license to publish or reproduce the published form of this article or allow others to do so, for United States Government purposes.

## Author Biography

*K. Aditya Mohan (ORCID: https://orcid.org/0000-0002-0921-6559; Email: mohan3@llnl.gov) received his B.Tech. degree in electronics and communication engineering from National Institute of Technology Karnataka in 2010. He received his M.S. and Ph.D. degrees in electrical and computer engineering from Purdue University in 2014 and 2017, respectively. He is currently a Signal and Image Processing Engineer in the Computational Engineering Division at Lawrence Livermore National Laboratory. His research interests include computational imaging, inverse problems, and machine learning.*

*Dula Parkinson (Email: dyparkinson@lbl.gov) received his BS in chemistry from Brigham Young University (2001), his PhD in physical chemistry from UC Berkeley (2006), and completed a postdoctoral fellowship at UC San Francisco (2010). Since then he has worked as a staff scientist at the Advanced Light Source at Lawrence Berkeley National Laboratory, focusing on instrumentation and image processing for synchrotron microCT.*

*Jefferson Cuadra (Email: cuadra1@llnl.gov) received his BS and PhD in mechanical engineering from the New Jersey Institute of Technology (2010) and from Drexel University (2015), respectively. He is part of the Materials Engineering Division at Lawrence Livermore National Laboratory in Livermore, CA. His work has been on metrology applications using X-ray CT, in addition to fatigue and fracture analysis of materials using FEM and nondestructive testing and evaluation.*